\documentclass[twocolumn, aps, prl,superscriptaddress,showpacs]{revtex4-1}

\usepackage{amsmath}
\usepackage{amssymb}
\usepackage{bm}
\usepackage{color}
\usepackage{graphicx}
\usepackage[latin1]{inputenc}
\usepackage[english]{babel}
\usepackage{tikz}
\newcommand*\circled[1]{\tikz[baseline=(char.base)]{
            \node[shape=circle,draw,inner sep=2pt] (char) {#1};}}
\begin{document}

% Use the \preprint command to place your local institutional report number
% on the title page in preprint mode.
% Multiple \preprint commands are allowed.
%\preprint{}

\title{Spectral Analysis of Topological Defects in an Artificial Spin-Ice Lattice}
%\title{Resonant dynamics in an artificial spin-ice lattice}
% something sexier like "Topological magnonics"?
%

\author{Sebastian Gliga\footnote{Present address: Max-Planck-Institut f\"{u}r Mikrostrukturphysik, Weinberg 2, D-06120 Halle(Saale), Germany.}}
\email{sgliga@mpi-halle.de}
\affiliation{Center for Nanoscale Materials, Argonne National Laboratory, Lemont, IL 60439, USA}
\affiliation{Max-Planck-Institut f\"{u}r Mikrostrukturphysik, Weinberg 2, D-06120 Halle, Germany}

\author{Attila K\'{a}kay}
\affiliation{Peter Gr\"{u}nberg Institut (PGI-6), Forschungszentrum J\"{u}lich GmbH, D-52428 J\"{u}lich, Germany}

\author{Riccardo Hertel}
\affiliation{Institut de Physique et Chimie des Mat\'{e}riaux de Strasbourg, Universit\'{e} de Strasbourg, CNRS UMR 7504, 67034 Strasbourg, France}

\author{Olle G. Heinonen}
\affiliation{Materials Science Division, Argonne National Laboratory, Lemont, IL 60439, USA}
\affiliation{Department of Physics and Astronomy, Northwestern University, 2145 Sheridan Rd., Evanston, IL 60208-3112}

%%%%%%%%%%%%%%%%%%%%%%%%%%%%%%%%%%%%%%%%%%%%%%%%%%%%%%%%%%%%%%
\begin{abstract}
\noindent Arrays of suitably patterned and arranged magnetic elements may display artificial spin-ice structures with topological defects in the magnetization, such as Dirac monopoles and Dirac strings. It is known that these defects 
strongly influence the quasi-static and equilibrium behavior of the spin-ice lattice. 
Here we study the eigenmode dynamics of such defects in a square lattice consisting of stadium-like thin film elements using micromagnetic simulations.  
We find that the topological defects display distinct signatures in the mode spectrum, providing a means to qualitatively and quantitatively analyze monopoles and  strings which can be measured experimentally. 
\end{abstract}

\maketitle
%%%%%%%%%%%%%%%%%%%%%%%%%%%%%%%%%%%%%%%%%%%%%%%%%%%%%%%%%%%%%%
%\begin{multicols}{2}

%\section{INTRODUCTION}
%\begin{figure}[b!]
%\includegraphics*[width=0.49\textwidth]{plate1-2}
%\caption{(Color online) Different vertex configurations in a square spin ice lattice. The first two ($RS$ and $DS$) are charge neutral while 
%$G$ and $G^*$ show two possible charged configurations. The arrows and colors represent the magnetization direction and the orientation 
%with respect to the vertex: orange ``{\it in}'' and blue ``{\it out}''. The magnetic charge is defined as positive when the net magnetization 
%points {\it out} of the vertex.
%}
%\label{fig:spin_ice_rules}
%\end{figure}

\noindent Spin ices are magnetic structures in which the interactions lead to frustration resulting in complex magnetic ordering and collective behavior~\cite{Morris2009,Jaubert2009,Fennell2009}.  
%Examples of such spin ice systems on the atomic scale are the pyrochlore structures 
%Dy$_2$Ti$_2$O$_7$ and Ho$_2$Ti$_2$O$_7$~\cite{Morris2009,Jaubert2009,Fennell2009}. 
%In these, the atomic spins arranged at the vertices of tetrahedra are forced by crystal fields to point either in or out of the tetrahedra.
{\it Artificial} spin-ice lattices are patterned nanoscale structures in which each element is small enough to have essentially uniform magnetization. The geometrical arrangement of the structures in the lattice leads to frustration by design. 
Examples of artificial spin ice are the Kagome lattice~\cite{Qi2008}
and the stadium square lattice, 
consisting of stadium-shaped elements on a square lattice~\cite{Wang2006,Moeller2006,Remhof2008}. 
%The magnetization at the vertices of the stadium square lattice satisfies simple spin-ice rules shown in
%Fig.~\ref{fig:spin_ice_rules}. 
The equilibrium configuration and magnetization reversal of such lattices have been the subject of intense 
study~\cite{Morgan2011,Morgan2011-2,Phatak2011,Phatak2012}. Topological defects, in which the magnetization at a vertex is in a one-in-three-out (or three-in-one-out) configuration~\cite{Gingras2009} have been observed in square lattices~\cite{Wang2006} and in Kagome lattices~\cite{Ladak2010}.
\begin{figure}[b]
\includegraphics*[width=0.45\textwidth]{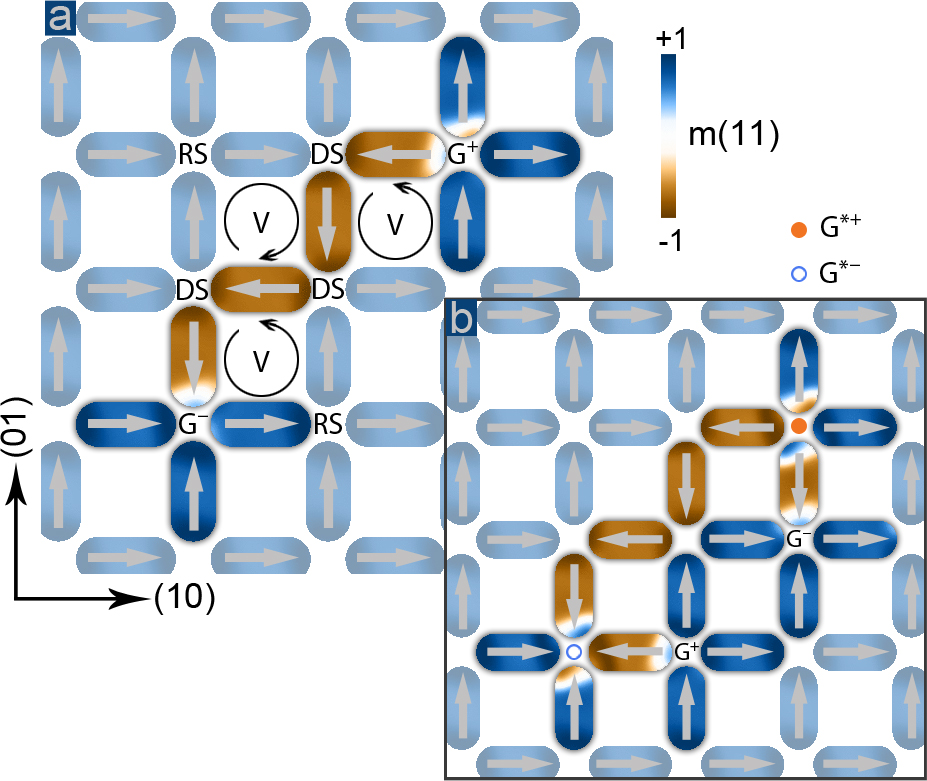}
\caption{(Color online) a) Simulated singly charged monopole pair ($G^+G^-$) linked by a Dirac string ($DS$). Vortices $V$ form along the string. The reference state ($ RS$) of the lattice is shown in pale blue, while the string and monopoles are highlighted. b) doubly charged monopoles can also
be observed experimentally, although relatively rarely~\cite{Wang2006}. In this case, each vertex with a doubly charged monopole leads to the formation of an adjacent oppositely-charged monopole (for example $G^-$ forms next to a $G^{*+}$ vertex). The colormap corresponds to the projection of the magnetization along the (11) direction of the string; the magnetic charge is taken as positive when the magnetization points out of the vertex.}
\label{fig:G_lattice}
\end{figure}
\begin{figure*}[ht!]
\includegraphics[width=0.9\textwidth]{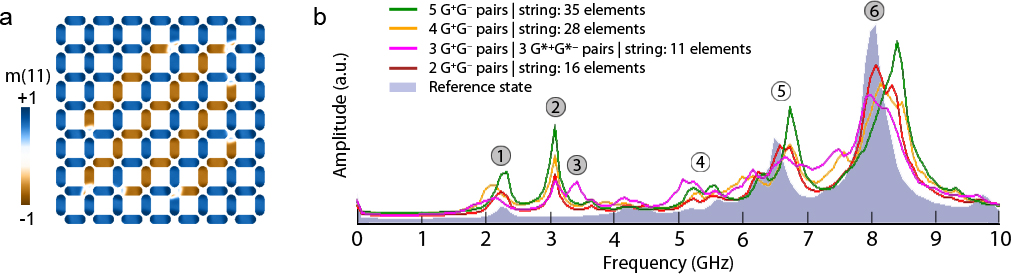}
\caption{(Color online) (a) Simulated lattice composed of 112 magnetic elements. It contains four monopole-antimonopole pairs connected by Dirac strings extending over 28 elements. 
(b) Evolution of the magnetization dynamics spectrum with increasing string length and number of monopole-antimonopole pairs compared to the reference state. The shaded labels correspond to the main, distinct signatures of the topological defects. The amplitude units are identical in all cases. 
}
\label{fig:spectra}
\end{figure*}
Such defects are similar to Dirac monopoles~\cite{Dirac1931} and typically occur in pairs ({\it e.g.} monopole-antimonopole) 
connected by a defect string, or ``Dirac string'', along which the magnetization is reversed with respect to a 
lattice reference state. Dirac strings have been observed in Kagome lattices~\cite{Mengotti2010} and square lattices~\cite{Phatak2011}.
(Fig.~\ref{fig:G_lattice}).  
Recent studies have shown that these defects profoundly affect the equilibrium behavior 
and the magnetization reversal of such 
lattices~\cite{Westphalen2008,Morgan2011,Kapaklis2012,Phatak2011,Phatak2012,Shen2012,Budrikis2012}. 
An interesting, yet unexplored question is whether the presence of such defects affects the resonant dynamics of the spin-ice lattice. 
Here, we investigate the frequency spectrum of eigenmodes, or resonances, of topological defects in a square spin ice lattice
using micromagnetic simulations. We show that each topological defect displays distinct and localized features, both spatially as well as in frequency. These features act as a fingerprint for the type of defect and moreover allow to quantify the number of defects present in the lattice.  
We also show how spectral features in a lattice arise from the magnetostatic coupling between the edge-like modes of isolated elements. 
The coupling leads to multiplets that are split depending on the local magnetization configuration.
In addition, the fundamental uniform mode of a single stadium is split in the presence of topological defects. 

\noindent We have studied the resonances of square spin ice lattices consisting of 112 Permalloy (Py) elements (Fig.~\ref{fig:spectra}a) with micromagnetic simulations using a fully three-dimensional finite-element algorithm based on the Landau-Lifshitz-Gilbert equation~\cite{Hertel07}. 
Each element (stadium) is 290 nm long, 130 nm wide, 20 nm thick and is discretized into about 22800 tetrahedral elements, with an average cell size corresponding to an equivalent cube with 3 nm side length.
%corresponding to a cell size of about 3 nm$^3$. 
The lattice constant of the array is 390 nm. The material parameters chosen for Py are the following: $A=13$ pJ/m (exchange constant), $\mu_0 M_s=1.0$ T, and zero magnetocrystalline anisotropy~\footnote{Similar results were also obtained using a fast Fourier transform method on a cubic lattice, with a mesh size of $\alt1.25$ nm, in order to account for the magnetostatic effects near the curved boundaries of the stadiums with high accuracy [R. E. Camley, B.V. McGrath, Y. Khivintsev, Z. Celinski, R. Adam, C. M. Schneider, and M. Grimsditch, Phys. Rev. B {\bf 78}, 024425 (2008)].}. 
We prepared a reference state (RS) of the lattice~\cite{Mol-2009} by saturating the array along the (11) direction
and adiabatically removing the external field (see Fig.~\ref{fig:G_lattice}). The system then relaxes into a state with the magnetization oriented along the $+y$ direction in the vertical elements and along the $+x$ direction for the horizontal elements of the array. 
We arbitrarily insert topological defects with charges $\pm 1$ and $\pm 2$ ($G^{\pm}$ and $G^{*\pm}$, 
respectively) %~\footnote{The notation G* emphasizes that doubly charged monopoles not only have a double magnetic charge, 
%but also a different spectral signature. They can be seen as an excited state, different from the singly charged monopole 
%[H. Abramowicz, {\it et al.} (ZEUS Collaboration), arXiv:1208.4468 (2012)].}
connected by Dirac strings ($DS$) by reversing the magnetization in selected stadiums, and subsequently relaxing the magnetization (see Fig.~\ref{fig:G_lattice}). These string defects correspond to typical observed configurations. The dynamics and spectrum of eigenmodes of the lattice are then obtained by applying a short perturbation (5 mT field pulse in the (11) direction for 20 ps) and integrating the Landau-Lifshitz-Gilbert equation in time. 

\noindent The excitation spectrum of the RS is shown in Fig.~\ref{fig:spectra}(b) as the grey, filled area. It is characterized by a few prominent peaks corresponding to distinct eigenmodes. The first one (at position \circled{1}), slightly above 2~GHz, corresponds to edge excitations of the outermost stadium elements at the boundary of the lattice. A weaker mode appears above 4~GHz corresponding to edge excitations of the magnetostatically coupled elements inside the lattice. 
Two modes, labeled \circled{4}, are visible slightly above 5 GHz, in which the elements on the boundary of the lattice display higher-order edge modes that couple to modes in elements adjacent to these boundary elements.
Above 6 GHz, label~\circled{5} indicates a mode corresponding to higher-order edge oscillations of the individual elements. This mode is split: owing to their smaller magnetostatic coupling, 
the elements at the boundary of the lattice oscillate at a slightly lower frequency than elements in the interior of the lattice. 
Finally, a mode at 8 GHz (position~\circled{6}) corresponds to the ferromagnetic resonance in which the oscillation is approximately uniform in the interior of the elements. 
The curves in Fig.~\ref{fig:spectra}(b) depict the spectrum evolution as topological defects and Dirac strings are inserted
into the lattice. 
%The RS spectrum is clearly modified with some mode frequencies shifting, and new peaks appearing. 
Near mode \circled{1} of the RS a new mode associated with oscillations of the $G^+G^-$ pairs appears. 
The amplitude of this mode increases with the number of monopole-antimonopole pairs. Later, we show that this mode is 
close in frequency to the edge modes of the RS, but is spatially localized at the monopoles. Therefore, it can
be distinguished from the edge modes either in the thermodynamic limit or by using detection techniques sensitive
to the interior of the lattice~\footnote{For example, time-resolved Kerr microscopy. In a lattice of macroscopic dimensions, a meander transmission line can be
placed near the interior of the lattice or patterned nano-elements such as magnetic tunnel junctions can be used to manipulate
and detect local magnetization through the spin torque effect.}.  
Similarly, around 5 GHz, a new mode associated with doubly charged monopoles unfolds alongside the RS mode at \circled{4}.  Above 6 GHz, oscillations of the monopole-antimonopole pairs give rise to a new mode within the lattice (at  position \circled{5}). When a large number of pairs is present, this mode becomes dominant. 
Finally, at \circled{6} the ferromagnetic resonance peak is split and shifted in the presence of Dirac strings. In addition to the modifications in the GS spectrum, two new resonances are observed: mode \circled{2} at 3 GHz corresponds to edge modes of elements in the interior of a Dirac string  while mode \circled{3} at 3.5~GHz is due to oscillations of doubly charged monopoles. 

\begin{figure}
\includegraphics*[width=0.47\textwidth]{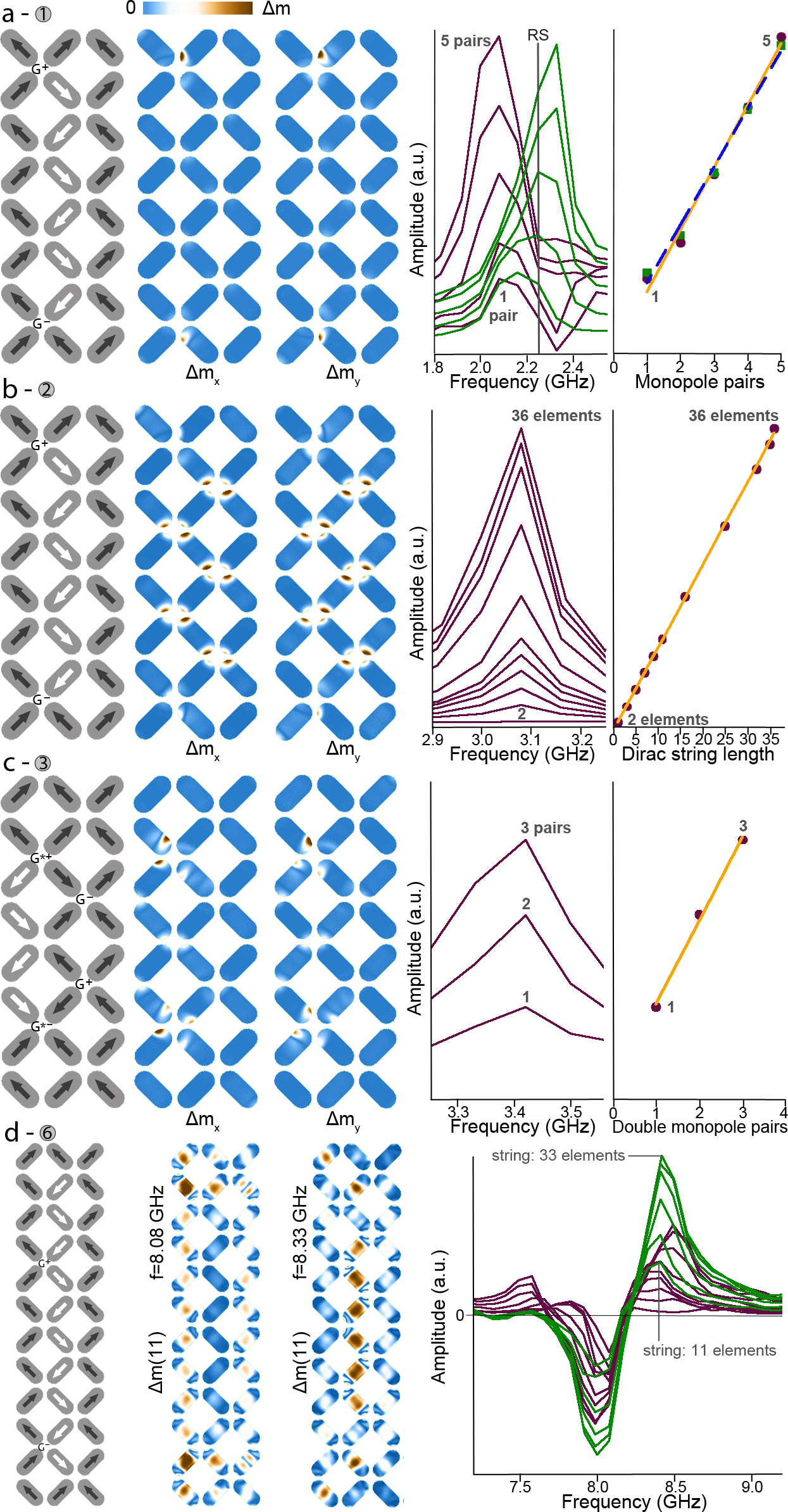}
\caption{(Color online) Examples of local oscillations of the magnetization ($\Delta m$) characteristic for monopoles (a), Dirac strings (b) 
and doubly charged monopoles (c). 
The first column shows the static magnetic configuration. The second and third columns respectively show the oscillations of the $x$ and $y$ components of the magnetization. The plots show the change in amplitude of the spectral signature of the different defects relative to the reference state. 
The last row (d) illustrates the splitting of the ferromagnetic resonance peak in the presence of a Dirac string. 
Labels are from Fig.~\ref{fig:spectra}.
}
\label{fig:topo_modes}
\end{figure}

\noindent We now investigate in detail the modes associated with the different topological defects. 
Figure~\ref{fig:topo_modes}a refers to the mode associated with a $G^+G^-$ pair at position~\circled{1} in Fig.~\ref{fig:spectra}. 
The static magnetization configuration is indicated by the arrows in the first column. The color maps (columns two and three) depict the 
oscillations of the $x$ and $y$ components of the magnetization, respectively, which are clearly localized at the monopoles. 
A splitting in the frequency peak is associated with the $G^+G^-$ resonance (graph, fourth column): 
Monopoles located on the boundary of the lattice (green curves) contribute to the amplitude of edge oscillations of the RS configuration, 
thereby increasing the amplitude of the RS mode (indicated in the figure) slightly above 2.2 GHz. 
Monopole configurations in the interior of the lattice oscillate at a lower frequency, around 2.1 GHz in this case (maroon curves).  The peak amplitude increases linearly with the number of monopole pairs in both cases as shown in the last column (simulations were carried-out for a maximum number of five pairs). 
It is therefore possible not only to identify the presence of monopoles in the lattice, but equally to {\it quantify} the number of pairs, given a reference point. 
Figure~\ref{fig:topo_modes}b refers to the oscillations of the stadiums in the Dirac string connecting a $G^+G^-$ pair at position~\circled{2} in Fig.~\ref{fig:spectra}. 
The spectral amplitude of this peak increases linearly with the total number of vertices in the string (string length). 
Figure~\ref{fig:topo_modes}c (position~\circled{3} in Fig.~\ref{fig:spectra}) shows the oscillations of a doubly charged monopole pair. 
Again, we note the linear relationship between the number of monopole pairs and the spectral amplitude of the mode. 
From this linear relationship, it can be concluded that the different strings and monopoles are decoupled from each other, even though they are rather closely packed (see Fig.~\ref{fig:spectra}a)~\footnote{When strings are adjacent to each other, they end in doubly charged monopoles rather than singly charged monopoles. Even then, however, we find that the strings are decoupled from each other. Stabilizing a doubly charged monopole pair therefore either requires the presence of a singly charged monopole-antimonopole pair (as shown in Fig. 1(b) or of two strings.}. Finally, Fig.~\ref{fig:topo_modes}d (position~\circled{6} in Fig.~\ref{fig:spectra} illustrates the splitting of the ferromagnetic resonance (FMR) mode~\footnote{%Direct diagonalization shows 
This peak actually corresponds to three near-degenerate distinct modes, two edge modes with two nodal lines (one odd
and one even), and the bulk-like FMR mode.}. The spatial oscillations are shown here by projecting the magnetization along the $(11)$ direction of the string, as in Fig.~\ref{fig:G_lattice}, in order to optimize the contrast. 
The two color maps correspond to spatial oscillations at the bulk FMR frequency of 8 GHz (second column) and at 8.3 GHz (third column), showing that the splitting stems from the Dirac string, which oscillates at a higher frequency.  The peak amplitude of the mode at 8.3 GHz therefore increases with increasing string length at the expense of the FMR peak at 8 GHz, as shown in the amplitude plot (rightmost panel). 
Again, a slight splitting occurs when the strings terminate on elements on the boundary of the lattice (green curves) or on elements in the 
interior lattice (maroon curves) 
because of the difference in magnetostatic coupling at the string extremities.  

\begin{figure}[t]
\includegraphics*[width=0.48\textwidth]{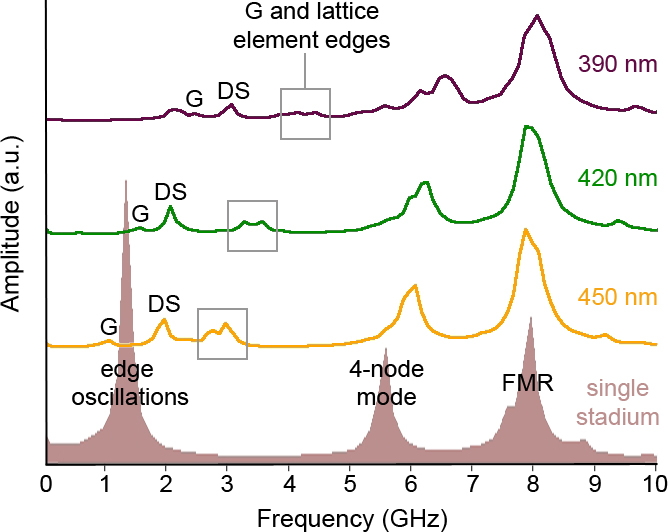}
\caption{(Color online) Evolution of the spectral features with decreasing element separation for an array containing one monopole-antimonopole pair connected by a string of six elements, compared to a single stadium. As the lattice constant of the array is reduced, the spectral features are blueshifted while the amplitude of the ferromagnetic resonance peak decreases. $G$: peak characteristic of the monopole pair oscillations. $DS$: peak characteristic for the Dirac string oscillations.  
}
\label{fig:lattice_spectrum}
\end{figure}

\noindent The magnetostatic coupling at the vertices drives the spectral features of the monopoles and the Dirac strings, and
the magnitude of this coupling depends on the lattice constant. The experimental observability of these features clearly 
depends on their robustness with respect to variations in the lattice constant. 
Figure~\ref{fig:lattice_spectrum} shows simulations with different lattice constants for a given configuration composed of one monopole-antimonopole pair linked by a string spanning six stadiums. 
For comparison, the spectrum of a single stadium with initial equilibrium magnetization
along the long axis of the element is shown (full shaded graph). 
There are three main peaks below 10 GHz for the single stadium. 
The first peak at 1.4 GHz corresponds to two modes of even and odd symmetry in which the dynamical part of the magnetization is localized 
near the edges of the stadium. 
Similarly, the peak at 5.7 GHz corresponds to 
two degenerate higher-order edge modes (one even and one odd) with additional nodal lines near the edges. 
Linear combinations of these edge modes couple magnetostatically in a lattice and give rise to multiplets, the splitting of which depends on the equilibrium structure of the magnetization at each vertex and on the 
lattice constant. Some of these linear combinations are localized near the vertices and are sensitive to the local
magnetization, and therefore to topological defects near the vertices. 
Finally, the peak at 8 GHz corresponds to the (bulk-like) uniform ferromagnetic resonance mode. 
In the presence of topological defects, with low magnetostatic coupling (lattice constant of 450 nm) the original peak at 1.4 GHz vanishes as the frequency of the edge oscillations is blueshifted to 2.8 GHz (indicated by the square frame in the figure). 
Given that monopole modes are connected to oscillations of the element edges, a peak associated to the monopole pair also appears at 3 GHz. At lower frequencies (1 GHz), another distinct peak corresponds to oscillations of the monopole pair. This peak is blueshifted when the lattice constant is decreased, and its frequency becomes comparable to the frequency of the edge excitation at elements on the lattice boundary at a spacing of 390 nm, as described earlier (Fig.~\ref{fig:spectra}~\circled{1}). An equally distinct peak corresponding to the Dirac string oscillation appears at 2 GHz for a 450 nm lattice constant and is blueshifted. The higher order edge mode found in a single element (at 5.7 GHz) is robust: it persists for every lattice separation, eventually splitting as described earlier. As expected, the frequency of the ferromagnetic resonance mode remains unchanged. It is interesting to note that even for a lattice constant of 450 nm, the elements within a lattice remain sufficiently coupled to display features essentially different from the single stadium spectrum. 

\noindent In summary, the presence and nature of topological defects
display a one-to-one correspondence with the spectral features, which thus provide a means of experimentally identifying the presence of the different defects. 
The frequencies of the modes lie well within the range of experimental detection techniques such as a broadband meanderline approach~\cite{Tsai-2009}, with an AC field provided by a microwave synthesizer and AC coils, or a pulsed field provided by a microwave stripline. Starting in the RS, applying a static reversal field with a magnitude close to the coercive field of the lattice should generate enough defects, Dirac monopoles and strings to provide sufficient signal to noise~\cite{Phatak2012}. This would also allow for an estimation of the number of defects as the reversal field strength is varied past the coercive field. 
Moreover, the spectral amplitude of these features is proportional to the number of defects present, enabling a quantitative analysis based on the spectra. 
The appearance of localized modes at  topological defects enables new possibilities of manipulating  magnetization modes in spin-ice lattices, for example by inserting monopole pairs using spin polarized currents and propagating Dirac strings through the lattice. This may allow emerging applications combining magnonics, in which  propagating spin waves in patterned magnetic thin films are manipulated, with artificial spin-ice lattices, in which localized spectral features based on topological defects can be tailored. 

%\section*{ACKNOWLEDGEMENTS}
\noindent SG wishes to thank Michael Sternberg for his invaluable support with the High-Performance Computing Cluster at the Center for Nanoscale Materials and Christine Stein for her advice with the graphics in this manuscript. 
This work was performed, in part, at the Center for Nanoscale Materials, a U.S. Department of Energy, Office of Science, Office of Basic Energy Sciences User Facility under Contract No. DE-AC02-06CH11357.

\bibliography{bibfile}

\end{document}